\documentclass[11pt,a4paper]{article} 
\usepackage[dvips]{graphicx}
\begin{document}
\thispagestyle{empty}
\renewcommand{\baselinestretch}{1.2}
\small\normalsize
\frenchspacing
\noindent
{\Large \textbf{Simultaneity in wavepacket  reduction}}
\\
\\
{\bf{Arthur Jabs}}
\renewcommand{\baselinestretch}{1}
\small\normalsize
\\
\\
Alumnus, Technical University Berlin. 

\noindent
Vo\ss str. 9, 10117 Berlin, Germany

\noindent
arthur.jabs@alumni.tu-berlin.de
\\
\\
(11 June 2015)
\newcommand{\rmi}{\mathrm{i}}
\vspace{15pt}

\noindent
{\bf{Abstract.}}  It is conjectured that the effects of reduction, which occur with the individual wavepackets in a system of two separated though entangled packets, occur simultaneously in the center frame (properly defined) of the entangled packets. Existing experiments are compatible with this type of nonlocality. Further experimental tests are suggested.

\begin{list}
{\textbf{Keywords:}}
{\setlength{\labelwidth}{2.0cm}
 \setlength{\leftmargin}{2.2cm}
 \setlength{\labelsep}{0.2cm} }
\item
entanglement, reduction, simultaneity, preferred frame, spooky action at a distance 
\end{list}

\newcommand{\rmf}{\mathrm{f}}
\newcommand{\rmd}{\mathrm{d}} 
\newcommand{\sbl}{\hspace{1pt}}
\newcommand{\bfitp}{\emph{\boldmath $p$}}
\newcommand{\bfitr}{\emph{\boldmath $r$}}
\newcommand{\bfitsr}{\emph{\footnotesize{\boldmath $r$}}}
\newcommand{\bfitQ}{\emph{\boldmath $Q$}}
\newcommand{\bfitk}{\emph{\boldmath $k$}}
\newcommand{\PSI}[1]{\Psi_{\textrm{\footnotesize{#1}}}}
\newcommand{\PHI}[1]{\Phi_{\textrm{\footnotesize{#1}}}}
\newcommand{\spsi}[1]{\psi_{\textrm{\footnotesize{#1}}}}
\newcommand{\pcop}{P_{\textrm{\footnotesize{Cop}}}}
\newcommand{\pred}{P_{\textrm{\footnotesize{red}}}}
\newcommand{\psis}{\psi_{\rm s}(\bfitr,t)}

\vspace{15pt}

That the effects of reduction (collapse) on two spatially separated but entangled wavepackets occur at spacelike distances in the laboratory frame has been shown in many experiments with two photon packets. This is the quantum mechanical nonlocality. The experiments   [1] - [4]   moreover showed that any hypothetical influence of one reduction effect on the other in some other frame would have to proceed at a speed that exceeds the speed of light $c$ by at least a factor of about $10^4$.

What is the exact value of that speed? At present there is no answer to that question. An interesting proposal is that the speed is infinite, in other words: that there is a reference frame where the two reduction effects occur simultaneously [5]. Indeed several candidates have been considered in the literature. In [7] the lab frame, the frame of the massive device that triggers the reduction, and the cosmic background radiation frame were considered, without  definite conclusion however. As quantum nonlocality anyway implies some violation of special relativity theory a rather drastic proposal is to replace the Lorentz transformation by the Tangherlini transformation [8]. Then simultaneity becomes frame independent, and the question of a preferred reference frame becomes meaningless.

In this note another proposal is advanced, which is motivated by the concept that a system of entangled wavepackets (an entangled system, for short) is a fundamental region of space [6, Secs. 2.3, 3.1]: simultaneity occurs in the `center frame' S$_{\rm C}$ of the entangled wavepackets and is restricted to the interior of the entangled system. The origin of this frame is defined as the mean position, with a wave function which describes the entangled system, for example: 
\begin{equation}
\bfitr_{\rm C}:=  \langle\bfitr(t)\rangle={\textstyle{\frac{1}{2}}} \int \bfitr\,\Big |\,\spsi{1}(\bfitr,a,t)\,   \spsi{2}(\bfitr,a,t) \pm    \spsi{1}(\bfitr,b,t) \, \spsi{2}(\bfitr,b,t)\,  \Big |^2 \,\rmd^3r.
\end{equation}

If in the center frame S$_{\rm C}$ the two reduction effects are connected by a
hypothetical velocity $\bf u_{\rm C}$ forming an angle $\alpha_{\rm C}$ with the relative velocity $\bf v$ between  S$_{\rm C}$ and the lab frame S$_{\rm L}$, then within the lab frame the hypothetical velocity $\bf u_{\rm L}$ forming an angle $\alpha_{\rm L}$ with $\bf v$ is [9]:
\begin{equation}
u_{\rm L}^2=\frac{u_{\rm C}^2+v^2+2u_{\rm C}v\cos\alpha_{\rm C} - [(u_{\rm C}v/c)\sin\alpha_{\rm C}]^2}  {[1+(u_{\rm C}v/c^2)\cos\alpha_{\rm C}]^2}
\end{equation}
and
\begin{equation}
\tan\alpha_{\rm L}=\tan\alpha_{\rm C}\,\frac{\sqrt{1-(v/c)^2}}{1+v/(u_{\rm C}\cos\alpha_{\rm C})}.
\end{equation}
Simultaneity in the center frame means $u_{\rm C}=\infty$, and formulas (2) and (3) turn into
\begin{equation}
u_{\rm L}^2=\frac{c^4}{v^2} \frac{1-[(v/c)\sin\alpha_{\rm C}]^2}{\cos^2\alpha_{\rm C}}
\end{equation}
\begin{equation}
\hspace{70pt}\tan\alpha_{\rm L}=\tan\alpha_{\rm C}\sqrt{1-(v/c)^2},    \hspace{30pt}(v\neq c).
\end{equation}
The inverse formulas are obtained by replacing $v$ by $-v$.
\vspace{4pt}

Some special cases are: 
\vspace{4pt}

(i) $\;\alpha_{\rm C}=0^{\circ} \hspace{38pt}   \Longrightarrow u_{\rm L}=c^2/v, \;\alpha_{\rm L}=0^{\circ}$,\hspace{40pt}de Broglie waves (`waves

 \hspace{255pt}of simultaneity' [10]).

(ii) $\;v=c  \hspace{50pt}  \Longrightarrow\;  u_{\rm L}=c$, 

(iii) $v=0  \hspace{50pt}  \Longrightarrow \;   u_{\rm L}=\infty, \; \alpha_{\rm L}=\alpha_{\rm C}$,

(iv) $v\neq c, \;\alpha_{\rm C}=90^{\circ}  \Longrightarrow \;   u_{\rm L}=\infty,  \; \alpha_{\rm L}=90^{\circ} $, \hspace{33pt}(`transverse simultaneity').

\vspace{6pt}

The hypothetical velocity $u_{\rm L}$ in the lab frame lies between $c$ and $\infty$, depending on the values of $v$ and $\alpha_{\rm C}$. The values of $v$ and $\alpha_{\rm L}$   are determined by the direction of each of the two wavepackets at the moment of the first reduction effect in the lab frame.
In the actual experiments [1 - 4] $u_{\rm L}>c$ is confirmed, and there is no indication of an upper limit to it. In [3] it was $v=0$, so that (iii) yields indeed $u_{\rm L}=\infty$. In [1, 2, 4] the angle $\alpha_{\rm L}$ is $90^{\circ} $  due to the symmetric arrangement of the receivers with respect to the source. Thus according to (iv)   $ \alpha_{\rm C}$ is also $90^{\circ} $  and $u_{\rm L}=\infty$.

Though the  experiments do not contradict the present proposal, the available data do not suffice to definitely confirm it. If it is possible to determine   $u_{\rm L}$,  one way to achieve a confirmation would be to calculate $u_{\rm L}(\alpha_{\rm L})$ by (4) and (5) as a function of  $v$ and $\alpha_{\rm C}$ and to compare this kind of dependence with the measured values.

\vspace{20pt} 

\noindent
{\textbf{Notes and References}} 
\begin{enumerate}
\renewcommand{\labelenumi}{[\arabic{enumi}]}
\hyphenpenalty=1000

\item Zbinden, H., Brendel, J., Tittel, W. et al.: Experimental Test of Relativistic Quantum State Collapse with Moving Reference Frames, arXiv:quant-ph/0002031 (J. Phys. A: Math. Gen. {\bf34}, 7103-7109 (2001))

\item Salart, D., Baas, A., Branciard, C. et al.: Testing spooky action at a distance, arXiv:0808.3316 (Nature {\bf454}, 861-864 (2008))

\item Cocciaro, B., Faetti, S. and Fronzoni, L.:  A lower bound for the velocity of quantum communications in the preferred frame, arXiv:1006.2697 (Phys. Lett. A {\bf375}, 379-384   (2011))

\item Yin, J., Cao, Y., Yong, H.-L. et al.: Bounding the speed of `spooky action at a distance', arXiv:1303.0614 (Phys. Rev. Lett. {\bf110}, 260407 (2013))

\item We prefer to speak of two effects of one reduction rather than of two reductions. The effects themselves may take some small time intervals; this is  considered negligible here.  The reduction is conceived to occur independently of being observed or not [6, Sec.~4]

\item Jabs, A.: Quantum mechanics in terms of realism, arXiv:quant-ph/9606017

\item Gisin, N., Scarani, V., Tittel, W. et al.: Optical tests of quantum nonlocality: from EPR-Bell tests towards experiments with moving observers, Ann. Phys. (Leipzig) {\bf9} (11-12), 831-841 (2000)

\item Tangherlini, F.R.: Galilean-Like  Transformation  Allowed  by  Gen\-er\-al Covar\-iance  and  Con\-sist\-ent  with  Special Relativity, http://dx.doi.org/10.4236/jmp.2014.55033 (Journal of Modern Physics {\bf5}, 230-243 (2014))

\item Pauli, W.: Theory of Relativity (Dover Publications, New York, 1958) p. 16

\item Rindler, W.: Introduction to Special Relativity, 2nd. ed. (Clarendon Press, Oxford, 1991) p. 83

\end{enumerate}
\bigskip
\hspace{5cm}
------------------------------
\end{document}